\documentclass[11pt,a4paper]{article}
\usepackage{pdflscape}
\usepackage{authblk,graphicx,rotating,amssymb,amsfonts,amsmath,color,rotating,multirow,array,setspace,pdfpages,lineno,natbib}
\usepackage[percent]{overpic}
\usepackage[margin=2.5cm]{geometry}
\usepackage[bookmarksopen=true,											
bookmarksnumbered=true,                      
plainpages=false,
pdfpagelabels,
colorlinks=false,                              
linkcolor=black, citecolor=black, urlcolor=cyan $•$
]{hyperref}

\title{Advancing the analysis of resilience of global phosphate flows} 

\author[a,b,c,*]{Matthias Raddant}
\author[d,e]{Martin Bertau}
\author[b,c]{Gerald Steiner}

\affil[a]{Graz University of Technology, Institute of Software Engineering and Artificial Intelligence, Inffeldgasse 16b/II, 8010 Graz, Austria}
\affil[b]{University for Continuing Education Krems, Department for Knowledge and Communication Management, Dr.-Karl-Dorrek-Stra{\ss}e 30,
3500 Krems, Austria}
\affil[c]{Complexity Science Hub Vienna, Josefst{\"a}dter Stra{\ss}e 39, 1080 Vienna, Austria}
\affil[d]{Institute of Chemical Technology, Freiberg University of Mining and Technology, Leipziger Straße 29, 09599 Freiberg, Germany}
\affil[e]{Fraunhofer Institute for Ceramic Technologies and Systems IKTS, Fraunhofer Technology Center for High-Performance Materials THM, Am St.-Niclas-Schacht 13, 09599 Freiberg, Germany}
\affil[*]{raddant@tugraz.at}
\date{} 
\begin{document}
\onehalfspacing

  \maketitle
  
  \begin{abstract}
  This paper introduces a novel method for estimating material flows, with a focus on tracing phosphate flows from mining countries to those using phosphate in agricultural production. Our approach integrates data on phosphate rock extraction, fertilizer use, and international trade of phosphate-related products. A key advantage of this method is that it does not require detailed data on material concentrations, as these are indirectly estimated within the model. We demonstrate that our model can reconstruct country-level phosphate flow matrices with a high degree of accuracy, thereby enhancing traditional material flow analyses. This method bridges the gap between conventional material flow analysis and the economic analysis of resilience of national supply chains, and it is applicable not only to phosphorus but also to other resource flows. We show how the estimated flows can support country-specific assessments of supply security: while global phosphate flows appear moderately concentrated, country-level analyses reveal significant disparities in import dependencies and, in some cases, substantially higher supplier concentration.
  
\noindent \emph{keywords: mineral resources, material flow analysis, hidden trade of materials, supply chains}
\end{abstract}

\section{Introduction}
Phosphorus (P) is a critical element in fertilizer production and, by extension, global food supply. Because agricultural practices continuously remove phosphorus from the soil, a stable supply of phosphate-based fertilizers is essential for both food security and economic development. While efforts to close phosphorus cycles at local and global scales are ongoing, agriculture will remain heavily dependent on phosphate fertilizers for the foreseeable future \citep[see, e.g.,][]{sws25,chen_graed_prod}.

These fertilizers are primarily produced from mined phosphate rock—mainly sedimentary, with a smaller share from igneous sources. Production is concentrated in a few countries, creating a complex and potentially fragile trade network. This network is susceptible to global supply and demand shocks, with implications for international stability.

Although phosphate rock is not currently scarce at the global level, and future supply is expected to meet demand, this is contingent on largely unrestricted trade. However, given that the top four producers -- China, the United States, Morocco, and Russia -- are geopolitically influential, the risk of politically motivated trade disruptions cannot be ignored \citep[see also][]{usgs23,kauw_reserves,barb_shortage,SCHOLZ201311}. 

As a result, phosphate trade -- including phosphate rock, fertilizers, and related products -- has been widely studied for its implications in supply chain vulnerabilities and price volatility \citep[see, e.g.,][]{wang_trade,weber_trade,klim_ober,p_fmarkets,chen_p_net}. In parallel, phosphorus flows within ecosystems -- through biomass production and food trade -- are also significant \citep[see][]{nesme_trade,p_flows0,rs_sustp}, as are losses during mining, fertilizer production, and agricultural use, all pointing to opportunities for improving efficiency \citep[see][]{HUANG_ploss,st_efficiency,prud_fows,ott_recy,t_paths}.

Agricultural use remains the dominant driver of global phosphorus demand. Imbalances in fertilizer use across countries reveal opportunities not only for optimizing crop yields but also for mitigating environmental impacts such as freshwater pollution. Additionally, projected increases in food demand will intensify challenges around sustainable phosphorus management \citep{p_imbalance,foley_sol,NEDEL_food}. This has led to the development of footprint metrics that assess local phosphorus demand against globally sustainable levels, accounting for the full production chain \citep[see][]{steffen}.

This paper, however, focuses on the supply security of phosphorus used in fertilizers. Such an assessment of resilience includes economic and geopolitical aspects, and requires a slightly different focus with respect to how we analyze flows compared to a traditional material flow analysis. Assessing supply security requires detailed knowledge of flows at the country level, has to incorporate a wide range of pathways, and has to be able to identify ultimate sources of materials no matter what path they have taken. While trade data on phosphate-related products is widely available, conventional methods struggle to translate this information into precise material flow models with these characteristics. Part of the problem is that trade data typically refer to products rather than the materials they contain, and existing methods often rely on product-specific concentration values to estimate physical flows. However, such concentration data are frequently incomplete or unavailable, especially for diverse products.

To address this, we propose a new top-down method that estimates material flows without requiring detailed concentration data. Instead of calculating flows from known concentrations (a bottom-up approach), we start from known extraction volumes and fertilizer use, and estimate the material content of traded goods by fitting them within a global flow model. In this framework, the system boundaries are defined by phosphate rock mining and fertilizer application.

Our model allows us to estimate, for the first time, the global, trade-based flows of phosphorus prior to biomass production. It captures transfers of P -- whether in raw phosphate rock, intermediates like phosphoric acid, or final fertilizer products -- between countries for agricultural use. This approach enables a detailed reconstruction of the first round of global phosphorus flows, forming a foundation for more accurate assessments of supply security and system resilience. Furthermore, the method is generalizable and can be adapted to other resource flows beyond phosphorus.

\section{Materials and methods}

\subsection{Overview}

It is important to note that, at the global level, the most reliable component of trade data is the monetary value of traded goods. This is because, in most countries, trade data originates from customs declarations, where monetary values are essential for determining tariffs and duties. While quantity data is often included, it can be less reliable and is only systematically available since 2006 -- insufficient for the longitudinal analysis conducted in this study.

Traditional material flow analysis typically relies on physical quantity data combined with estimates of material concentrations to derive flows. Monetary-based flows are more common when multi-regional input-output data is used (MRIO). This allows the disaggregation of flows on a sectoral level -- but is for many applications, including the P flows before bio-mass production, not detailed enough \citep[see, e.g.,][]{mfa_grae,2flows}. In contrast, our method is designed to work directly with monetary trade values. Since we aim to estimate relative shares of global trade rather than absolute quantities, using value data is sufficient for our purposes, as detailed in Section~\ref{sec:flows}. Additionally, trade values often implicitly reflect product quality or concentration (e.g., higher-grade materials usually command higher prices), which can serve as a proxy for material content.

Trade data is reported at varying levels of detail. For cross-country comparability, we use the Harmonized System (HS) at the 6-digit level, which includes about 6,700 product categories. While this level ensures broad international comparability, individual product categories still contain a degree of heterogeneity.

Trade data is not without its challenges: gaps, inconsistencies, and valuation differences -- including those arising from exchange rate fluctuations -- are well-documented \citep[see, e.g.,][]{CHEN_data,murya_data}. To address these issues, we use a cleaned and reconciled version of the UN Comtrade database, published as Atlas \citep{atlas23}, where import and export values are harmonized and reported on an FOB (free on board) basis.

In our analysis, we approximate the flow of phosphorus between countries using annual aggregate trade values. To derive physical flows, we estimate the average phosphorus content per USD for each relevant product category. These estimates allow us to convert value-based flows into material flows. The total of these category-specific flows is constrained to match known global phosphate flows for each year, as defined by the boundaries of our system.

Our system boundaries are on the one side set by P production, approximated by phosphate rock mining data by country. On the other side they are set by the amount of P used in fertilizer application by country.
To align data across production, trade, and use, we normalize country-level values relative to total global values in each year. This produces a flow matrix 
$F_{ij}$ where each entry in row $i$ represents the fraction of global phosphorus mined in country $i$ that is ultimately used in country $j$.
Row sums of $F$ represent the share of global phosphorus mined in a given country.
Column sums represent the share of global phosphorus used in a given country.
The entire matrix is normalized such that all elements sum to 1.
To convert these shares into absolute quantities (e.g., in metric tons of $P_2O_5$), the values can be multiplied by the total global phosphate availability in each year (e.g., as reported by a geological survey). This yields a detailed country-by-country breakdown of the origins and destinations of phosphorus applied as fertilizer.

\subsection{Data}\label{sec:data}

\begin{figure}[htb]
  \tiny
\begin{center}
    \includegraphics[width=0.85\textwidth,trim = 50 290 50 270]{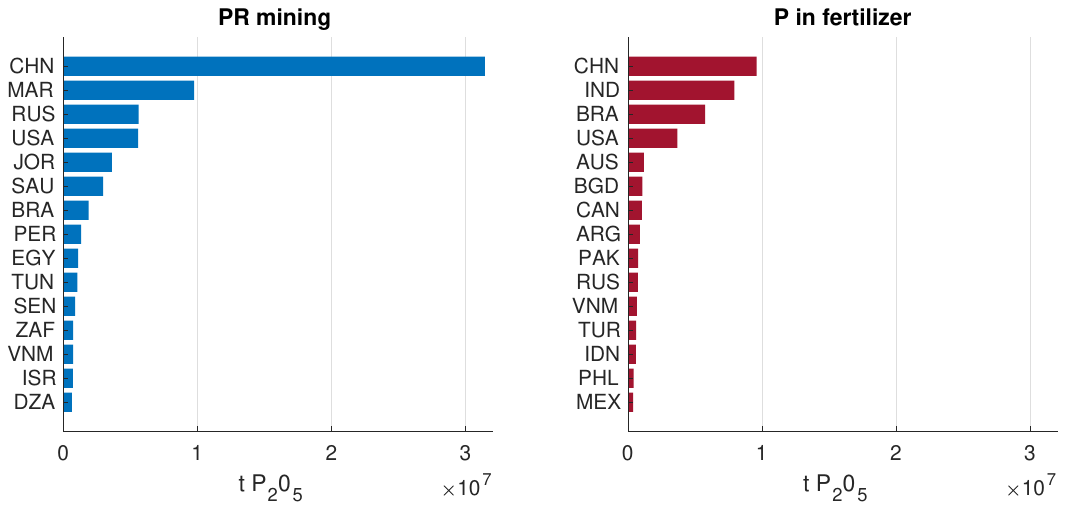}
  \end{center}
  \caption{World's largest producers of phosphate rock (left panel) and largest phosphoric fertilizer users (right panel) in 2022 according to World Mining data and the FAO.}\label{fig:produse}
\end{figure}

Data on the mining of phosphate rock (in the following abbreviated as PR) for the period from 2001--2022 was obtained from the World Mining database \citep{wmdata23}. This data set contains information on the global mining activity, in particular this refers to 39 countries with the current ability to mine PR, of which 37 reported mining in 2022.\footnote{We are aware that other data sources exist, especially for data on mining, yet we choose World Mining because this source is very comprehensive especially for smaller countries. For an overview of data sources on phosphate and the related problems  the reader is referred to \citet{p_fog}.} The total amount mined in 2022 was 71.1 million metr. t of $P_2O_5$, see the left panel of figure \ref{fig:produse}.

For the data on trade we utilize the Comtrade database \citep{comtrade23} in the version provided as Atlas \citep{atlas23}.  We use data from those product categories that are known to contain significant amounts of P (verified in discussions with industry experts). We utilize data on the trade of products classified (according to the HS classification) in the categories 251010,	251020,	280910,	280920,	283510,	283521,	283522,	283523,	283524,	283525,	283526,	283529,	283531,	283539,	310310,	310320,	310390,	310510,	310520,	310530,	310540,	310551,	310559,	310560 and	310590. These 25 categories can be hierarchically aggregated into five more general categories, namely natural calcium phosphate (phosphate rock), phosphoric acid, phosphates (mostly industrial use), phosphatic fertilizers and mixed fertilizers.

The database contains information on the amounts traded on a bilateral basis in USD for 234 countries. For example, in 2022 the total trade volume for all the goods in the categories mentioned above amounted to 62.1 billion USD. Of this 4.68 billion USD were traded in natural calcium phosphates, 7.25 billion USD were traded in phosphoric acid, 5.66 billion USD were traded in phosphates, 3.72 billion USD were traded in phosphatic fertilizers, and 40.82 billion USD were traded in mixed fertilizers. However, in this study we will consider a weighted sum of these categories (details in section \ref{sec:weights}). For an overview of global trade, as well as details about trade of China, the US, Russia, Morocco and India, the reader is referred to appendix \ref{sec:a}.

The use of phosphate can be approximated by data from \cite{fao_data23}. This data set contains information on the import, export, production and agricultural use of fertilizer in 167 countries (these countries are responsible for roughly 95\% of P-related trade). This data is typically available with a delay of 2 years, which determines that the current endpoint for our analysis is 2022. For example, the agricultural use of P in fertilizer for 2022 is reported at 41.9 million metr. t of $P_2O_5$, see also the right panel of figure \ref{fig:produse}. 
Hence, when one compares the figures of PR mining and P fertilizer use (and averages these over a few years) one could come to the conclusion that about 70\% of mined phosphate rock ends up as fertilizer. Considering losses in the production processes of fertilizer and tendential under-reporting in fertilizer use, we know that the actual share is in fact higher. Therefore, studies have looked at the technical processes that are involved in fertilizer production. For example, we know that about 90\% of processed mined phosphate is used in a chemical wet process and mostly converted to phosphoric acid, out of which about 82\% is used to produce fertilizer. Considering further that 15\% of P fertilizers are not made from phosphoric acid, we can approximate a lower bound of 80\% of total P used in fertilizer \citep[see][]{herman_processing}.  Other studies estimate a higher fraction of P fertilizer use, at the upper end of the spectrum \citet{fao04} estimates that 90\% of mined PR is used by the fertilizer industry.\footnote{These differences can partly be explained by slightly varying approximations of the shares of fertilizer production processes, as well as by differences in the accounting for animal feed supplements ($\sim$7\%).} 

In any case, we note that in our analysis we make the simplifying assumption that the entire PR production (and no other source) is used as fertilizer. This means that the P used for other purposes \citep[e.g. via phosphorus compounds, see also][]{shinh} is handled as if it flows in the same way. We also implicitly assume that mining and use take place in the same year, and we neglect the effect of changes in stocks.

\subsection{Trade-based flows}\label{sec:flows}

 \begin{figure}[tb]
  \tiny
\begin{center}
    \includegraphics[width=0.8\textwidth, trim = 35 180 10 160, clip=true]{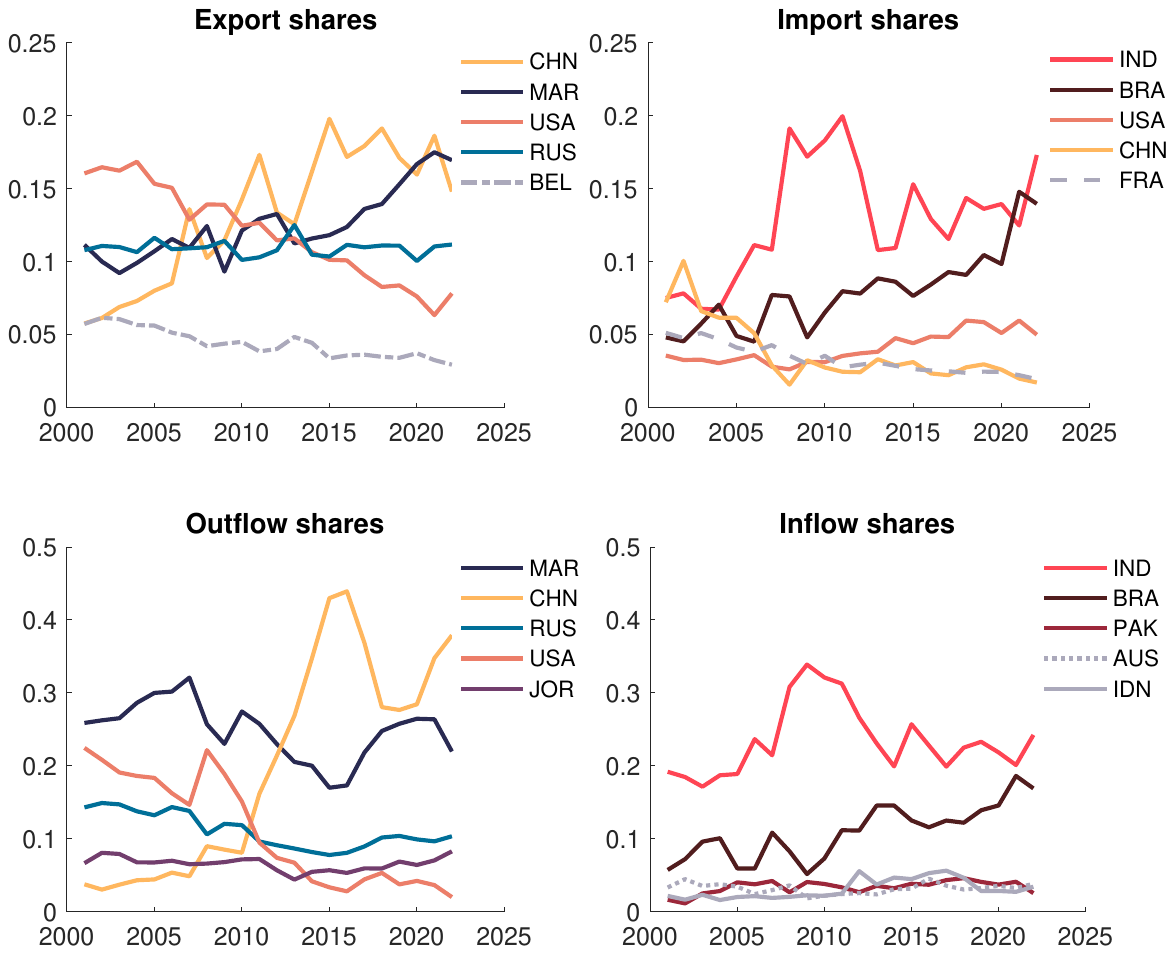}
  \end{center}
  \caption{Shares in P-related trade by country over time. The top panels show the five largest exporters and importers of P-related good. The United States and China are the only countries that appear in both panels. The bottom panels show the largest exports and importers in a material sense, as derived from our analysis ($F^{M5}$). The rise and fall back of the Chinese outflow share is the most noticeable change over time.}\label{fig:tradesh}
\end{figure}

The phosphate trade network reflects large global dissimilarities between the mining and the agricultural use of P. The shares in P-related trade over time are visualized in figure \ref{fig:tradesh}. China, Morocco and Russia are large exporters of P-related goods. India and Brazil are the largest importers. The position of the USA is unique because of the very high share of locally used P. Still, the USA is a net exporter, since it exports large amounts of fertilizer, which dominates its value-based P trade statistics. 

The trade network is however not sufficient to draw conclusions about the flow of P in a material sense (see the bottom panels of figure \ref{fig:tradesh}). One reason is that goods in the different product categories contain P in very different amounts per USD value.  The second reason is that countries can appear as net exporters of P while not mining any phosphate rock at all. This can happen because they act as hub for the trade and production of P fertilizer. Third, we must account for flows within countries, this concerns fertilizer that is used in the same country where phosphate rock is mined.

In the following we will show that these problems can however be overcome by transforming the matrix of imports and exports into a matrix of shares of net flows. We achieve this by applying some accounting identities and by re-evaluating indirect P-flows between countries where necessary.

First of all, let us start by defining two vectors that carry the information on PR mining and on the agricultural use of P-based fertilizer by country, $M^*$ and $U^{*}$. The data is taken from the World Mining and FAOSTAT database respectively. Both row vectors contain $N$ entries, where $N$ is the number of countries, in our case 234, plus one dummy for trades to undeclared places. For further calculations it will be useful to normalize these vectors such that they represent the shares in global mining and use receptively.\footnote{For the countries for which no data on fertilizer use is available we assume that their share is the same as their share of P net imports.}  Hence, we define 

\begin{equation}
M_i = \frac{M_i^*} {\sum_{i=1}^N M_i*}
\end{equation}
 and

\begin{equation}
 U_i = \frac{U_i^{*}} {\sum_{i=1}^N U^*_i} \; .
\end{equation} 

For each pair of countries, the data set contains the traded value in USD for the $C = 25$ product (goods) categories. These can be represented by separate matrices $G_{1 \hdots C}$ with dimensions $N \times N$. 
In these matrices each entry in row $i$ represents exports of country $i$ towards the country in column $j$. Since the imports and exports in the database have been harmonized, the matrices can equivalently be interpreted in a column-wise fashion, thus entries in column $j$ represent imports to country $j$ from country $i$.

To utilize the information on P trade to inform us about actual P flows it is useful to first consider only the net in P trade. For this we can first calculate a matrix of net exports by subtracting imports from exports for each pair of countries and by dropping negative values, hence

\begin{equation}\label{eq:net}
G^{net}_{i,j} = max \left( 0 \; , (G - G')_{i,j} \; \right) \; .
\end{equation}

To account for the fact that each of these categories represents goods with a specific average P content, a weighting for these categories is necessary. Hence, in the following we will work with matrices of weighted P trade $T^{net}$,

\begin{equation}\label{eq:trade}
	T^{net} = \sum_{c = 1}^C w_c G^{net}_c \; ,
\end{equation}   
where $w_c$ is the weight given to goods category $c$ and $\sum_c w_c = 1$. For the moment however, we assume equal weights for all goods categories. We will discuss optimized weighting schemes in section \ref{sec:weights}.

We further remove reciprocal flows between countries from $T^{net}$ through different goods categories by applying equation (\ref{eq:net}) equivalently. We then normalize the matrix of net exports by the total and obtain

\begin{equation}
T^{norm} = \frac{T^{net}}{\sum_{i=1}^N \sum_{j=1}^N T^{net}_{i,j}} \; .
\end{equation}

To derive a complete picture of trade-based flows we further must realize that we are missing information on P that is not traded but that stays within the country where it is mined. We can however approximate this data by assuming that exported P must originates from mined P that has not been used domestically or from net imports $Im_i$ from other countries.\footnote{The net imports can be calculated from $Im_i = \gamma_{tr} \sum_j T_{i,j}^{net}$, where $\gamma_{tr} = 0.6$ corrects for the overall share of traded vs locally used P that we can approximate from the aggregate data.} Hence, we can calculate a vector with the implied share of exported P, $\hat{E}$ as

\begin{equation}\label{eq:trade_exp}
\hat{E}_i = max \left( 0 \; , M_i - U_i + Im_i \right)  \; .
\end{equation}

From this we can calculate a vector $\hat{L}$ with implied shares of P that remain locally as

\begin{equation}\label{eq:self}
\hat{L} = M - \hat{E} \; .
\end{equation}

This gives us now the opportunity to include local P use as the diagonal entries into the normalized trade matrix, which (after re-scaling) results in the flow matrix $F^{M1}$, the first version of a P-flow representation. In this matrix each element represents the material share of P moved from country $i$ to country $j$. 

\begin{equation}\label{eq:net1}
	F^{M1} = T^{norm} \left( 1- \sum_{i=1}^N \hat{L}_i \right) + \hat{L} I \; ,
\end{equation}

where $I$ is a $N \times N$ identity matrix.

\subsection{Optimal category weights and fit}\label{sec:weights}

In order to validate if any flow matrix $F$ is a good approximation of the true material flows in the system, we have to evaluate if it resembles the distribution of PR mining and P used in agriculture correctly.\footnote{Note that a mathematically solution for an optimal flow  is $F^{opt}= M \; U'$, which would imply flows proportional to local use from all countries that mine PR. We are however interested in a flow matrix that resembles the actual material flows, which are distinct from $F^{opt}$ and for which trade flows are known to be a latent version.}
For this we will assume that the reported data on PR mining and the data on agricultural use of P represent the true shares of P origination and use, which of course is a simplification.
For the analysis of the fit, it makes sense to use the relationships between the matrix $F$ with the vectors representing the shares of P mining and use, $M$ and $U$.
We can derive a vector of predicted P use $\hat{U}$  by multiplying with the flow matrix

\begin{equation}\label{eq:pred1}
 \hat{U} = J \; F \; ,
\end{equation}

where $J$ is a vector of ones with dimensions $1 \times N$.
Similarly, we can predict the implied mining by using the transposed of $F$ and calculate

\begin{equation}\label{eq:pred2}
\hat{M} = J \; F'.
\end{equation}

These relationships can also be expressed as the row and column sums of $F$.\footnote{We note that equation \ref{eq:self} implies that the entries on the main diagonal of $F$ contain information on $U$ and $M$ by construction.}
We can use the predictions and compare $\hat{U}$ with $U$ and $\hat{M}$ with $M$ to evaluate the accuracy of $F$. We can in fact use this evaluation to solve one remaining problem in the calculation of $F$ in the first place. 

In particular, we have to investigate if the weighting scheme in the calculation of the trade matrix (see eq. \ref{eq:trade}) can be improved. In an ideal setting we would try to calculate the exact P content of each trade relationship that each country has with each other country for each goods category. This, however, is not feasible, due to resource constraints and data availability. We can however obtain the approximate P content indirectly by estimating the optimal weighting scheme that results in the flow matrix that gives us the most likely actual P flows between countries.

For this purpose we define a function $D$ that evaluates the difference between the estimated and the observed quantities in terms of a weighted sum of absolute errors.\footnote{Since we are interested in minimizing the mis-allocation of P-flows globally, we found that absolute errors are in this case a better choice than squared errors, since the latter would (due to the size distribution of countries) lead to an undesirable focus on mainly fitting few large countries' P data.}

\begin{equation}\label{eq:D}
 \emph{D} =  \alpha_U \sum_{i=1}^N \left| \hat{U_i} - U_i  \right|  + \alpha_M \sum_{i=1}^N \left| \hat{M}_i - M_i  \right|   
\end{equation}

We obtain optimal weights by employing a non-linear optimization of $D$ by choosing positive weights $c$ in equation \ref{eq:trade} such that $D$ is minimized. Repeating the calculations described in section \ref{sec:flows} generates an improved flow matrix $F^{M2}$.

We found that it is useful to give the mining part of equation \ref{eq:D} a slightly higher weight than the use part. This is caused by the fact that the fertilizer use data is rather noisy, because for many countries figures are estimated or approximated. In particular we choose $\alpha_M = \frac{2}{3}$ and $\alpha_U = \frac{1}{3}$.\footnote{We found that the exact weighting of $\alpha_{U,M}$ has only negligible effects on our results (see also section \ref{sec:only}). Values of $\alpha_U > 0.5$ will however lead to noisy estimation results.}  
Goods categories that contain only very few entries have been omitted to guarantee that an optimum for \ref{eq:D} can be found. We found that it is sufficient to include the 11 goods categories with the most volume (which account for 90 \% of the overall trade in the 25 P-related categories). While it is numerically possible to include up to around 16 goods categories in the estimation, this does not improve the overall fit of the estimated P flows.

\begin{table}[htb]
\centering
\begin{tabular}{c | c c c c c | c }
Model  & $F^{M1}$  & $F^{M2}$  & $F^{M3}$  & $F^{M4}$  & $F^{M5}$  & $F^{M6}$ \\
                       & \small{trade} & \small{opt. weights} & \small{corr. origin} & \small{corr. trade} & \small{corr. $M$} & \small{weights $\alpha_M$} \\
                       \hline
  & & & & & &                     \\
$ \langle R^2_{min} \rangle $ & 0.9229	&	0.9770	&	0.9835	&	0.9863	&	0.9977	&	0.9980	\\

 $ \langle R^2_{use} \rangle $	&0.9537	&	0.9758	&	0.9840	&	0.9866	&	0.9896	&	0.9762	\\

 $ \langle R^2 \rangle $ &  0.9383	&	0.9762	&	0.9837	&	0.9865	&	0.9937	&	0.9871	\\
 
$\pm$   & \small{0.0183}	&	\small{0.0079}	&	\small{0.0062}	&	\small{0.0057}	&	\small{0.0045}	&	\small{0.0053}	\\
	 & & & & & &                     \\
 $ \langle D \rangle $ & 0.3721	&	0.2092	&	0.1588	&	0.1493	&	0.0823	&	0.1079	\\

$\pm$    &  \small{0.0350}	&	\small{0.0273}	&	\small{0.0309}	&	\small{0.0287}	&	\small{0.0260}	&	\small{0.0236}	\\

\end{tabular}
\caption{Fit of the flow matrix with observed PR mining and P use. The table shows the average $R^2$ value (for the years 2001--2022) for the fit of $F$ with $M$ and $U$ separately as well as combined, as well as the standard deviation. Further we show the average $D$ as well as its standard deviation. Note that $F^{M2}$ to $F^{M6}$ have been calculated based on weights optimized for each model.}\label{tab:fit}
\end{table}

The difference between the flow matrices $F^{M1}$ and $F^{M2}$ is that the latter is now based more heavily on goods categories that are responsible for a high material P flow, irrespective of the monetary value. In the calculation of the flow, we now mainly rely on the information from the trade in phosphoric acid, DAP and MAP fertilizers, as well as calcium phosphate. To a lesser degree also superphosphates, sodium triphosphate, and other fertilizer forms are considered. The optimal weighting scheme shows some variation from year to year. For an example see figure \ref{fig:pie} in the appendix.

Table \ref{tab:fit} gives an overview of the fit. We note that in all our analysis we separately estimate the weights and calculate the flow matrix for the years 2001--2022.
While using net exports already produces a good fit of the flow matrix with the data, we observe that just by optimizing the weights we can improve the average $R^2$ from 94\% to 98\% (first vs. second column).
In the following we will investigate in how far further adjustments to the trade data can improve these results.

\subsection{Correcting for PR origination}

In our analysis we rely on the fact that overall, the recorded net trade must resemble the actual flow of P to a large amount. Some distortion by trade activity, i.e., the fact that P may travel through several countries and possibly several goods categories before being used, is inevitable. While we cannot correct for the entirety of these effects, it is possible to make noticeable improvements by using information on the mining activity in the different countries.

When we evaluate the row sums of $F^{M2}$, we observe that some countries show up as net exporters of P without having sufficient mining activity. To derive a flow matrix that is consistent with the qualitative aspects of the mining data this aspect must be corrected. This means that we have to shift these exports to the most likely actual origin, i.e., the countries that mine PR.

To achieve this, we perform the following algorithm. For each country $i$ that is not mining PR we consider its excess P exports $x_i$ given by entries in row $i$ in $F^{M2}$,

\begin{equation}
x_i  = \sum_{j=1}^N F^{M2}_{i,j}.
\end{equation}

 We then calculate the import shares of country $i$ from countries $j$ that do mine. We define the set of these countries as $j \in m$.
\begin{equation}
s_{m,i}^{imp} = \frac{F^{M2}_{m,i}}{\sum_m F_{m,i}^{M2}} \;,
\end{equation}

Equivalently, the export shares to all countries $j$ are given by

\begin{equation}\label{eq:excess}
s_{i,j}^{exp} = \frac{F_{i,j}^{M2}}{\sum_j F_{i,j}^{M2}} \;.
\end{equation}

Further we calculate the amount of imports into country $i$ that are used locally and are not exported, which is

\begin{equation}
 l_i = max(0 , \sum_{i=1}^N F_{i,j}^{M2}  - \sum_{i=1}^N F_{j,i}^{M2} ) \; .
\end{equation} 

In the flow matrix we then add the excess exports of country $i$ to the exports of countries that mine PR, given by $x_i (s_i^{imp} s_i^{exp})$. Here we make use of the fact that $s^{imp}$ is a row vector and $s^{exp}$ is a column vector. 
We then remove these exports from the export of the non-mining countries. Further we must scale down the imports from the mining countries to the excess exporter by multiplying these imports by $l_i (\sum_i F^{M2}_{i,j})^{-1}$, which means that we correct the imports to the amount that the country itself uses. We denote the resulting matrix as $F^{M2*}$. In total this procedure removes about 5\% of the flows accounted for in $F^{M2}$. Since this matrix was normalized (see eq. \ref{eq:net1}) we must re-instate this normalization after our correction algorithm. To achieve this and to derive at our improved flow matrix $F^{M3}$ we have to consider that we should only re-scale the trade (off-diagonal) part of $F^{M2*}$, since the entries on the main diagonal were calculated from sources which absolute value in terms of P was known. The normalization that takes this into account is given by

\begin{equation}\label{eq:rebal}
 F^{M3} = \frac{F^{M2*} - diag(F^{M2*})}{\sum_i \sum_j F_{ij}^{M2*}} \; \left( 1- tr(F^{M2}) \right) + diag(F^{M2}) \;.
\end{equation}

The statistics in table \ref{tab:fit} show that these approximate corrections do in fact improve the fit of the model significantly. Interestingly the fit does not only improve with the mining data but also with the use data, which indicates that the correction has also a positive indirect effect on the accuracy of the column sums of the flow matrix.

\subsection{Correcting for trade}

In the previous section we have shown how we correct the data for countries that do not mine phosphate rock. For these countries it was self-evident that they could not be the source of relevant phosphate flows. For countries that do mine PR a similar procedure can be used to approximately correct for trade flows where an intermediary is involved in a trade chain. The most relevant case here are probably the USA, which are both im- and exporter of phosphate, as well as producer. Until now our calculations have implicitly assumed that, for example, imports to the US have completely been used in the US and that all exports from the US originate from locally mined PR. A more realistic assumption is of course that imported P is both used domestically as well as re-exported (albeit likely in processed form). The share that is re-exported must be attributed to the original source country.

To analyze such trade chains, it is useful to look at a flow matrix with the domestic  markets removed, and start by defining a matrix $F^{D3} = F^{M3} - diag(F^{M3})$.
We will now look at all the countries with PR mining\footnote{Countries with less than 1\% of global PR production have been excluded from this correction.} and again label this set of countries as $m$. We can calculate how much P can \emph{potentially} flow through a country $i$ as

\begin{equation}
p_i = min \left( \sum_{j=1}^N F^{D3}_{ij} \; , \; \sum_{i=1}^N F^{D3}_{ij} \right) .
\end{equation}

The in- and outflows for a country $i$ in terms of shares of the total flow are given by the vectors

\begin{equation}
 f^{in}_i = \frac{F^{D3}_{m,i}}{\sum_{i=1}^N F^{D3}_{m,i}}
\end{equation}
and 
\begin{equation}
 f^{out}_i = \frac{F^{D3}_{i,j}}{\sum_{j=1}^N F^{D3}_{i,j}}.
\end{equation}

We define the share of the potential through-flow that we think should be re-accounted as $\gamma_p$. We can now correct the flows by adding the through-flow of each country under consideration to the outflow of the original exporter by the operation

\begin{equation}
 F^{M3*} =  F^{M3} + \gamma_p \; p_i \left( f_i^{in} f_i^{out}   \right) .
\end{equation}

We then have to reduce the outflow of country $i$ by $ \gamma_p \; p_i \; f_i^{out} $ and reduce the inflow by $ \gamma_p \; p_i \; f_i^{in} $. Re-normalization of the resulting matrix (see equation \ref{eq:rebal}) gives $F^{M4}$.

The parameter $\gamma_p$ can be approximated by evaluating the fit of the model for values $ 0 \leq \gamma_p \leq 1$. We found the optimum to be $0.4$, which means that countries that mine PR re-export on average 40\% of the P that they import additionally to their own production. We note that in reality longer chains of trade than considered here exist, however, we found that these are very likely not relevant for our results.\footnote{This has two reasons. The first is that our analysis is already based on shares of net exports. The second reason is that for longer chains to be relevant we would have to observe two intermediate countries with large trade volume in the middle of such a chain, which we do not. In fact, the only case where we do suspect longer chains of trade are smaller remote countries, for example in Africa, which do however have small P in-flow. Also, for such countries the trade data is rather inaccurate, which makes 2nd-order corrections impractical.} The statistics of the improvement in fit from this step are shown in table \ref{tab:fit}. The difference between $F^{M3}$ and $F^{M4}$ is in fact relatively small compared with the previous correction steps. 

\subsection{Scaling outflow to mining data}\label{sec:tie}

The last correction deals with the overall differences between the PR production data and the implied mining of $F^{M4}$. Assuming that the PR mining data overall has a good level of accuracy (and is less distorted by price variations than the trade data), we can scale the entries in the rows of the mining countries in $F^{M4}$ in such a way that their sum will closely match $M$. This operation is of course only useful as long as we do not negatively impact the column sums in terms of its prediction of fertilizer use. We note that this optional step in principle is an extension to the steps taken in section \ref{sec:flows}, with the difference that here we condition the entire row sums (and not only the diagonal elements) of the flow matrix on $M$.

Again, we define a partial flow matrix without the local use as $F^{D4} = F^{M4} - diag(F^{M4})$.
We then calculate how much P out-flow should be available in each country. This share is given by 
\begin{equation}
 \hat{f}_i^{out} = M_i - F^{M4}_{ii},
\end{equation}
i.e., the mining in country $i$ minus the local use.
We then calculate a correction factor $r_i$ based on the ratio of observed versus expected out-flow of country $i$, given by
\begin{equation}\label{eq:r}
 r_{i} = \left( \frac{\sum_{j=1}^N F_{ij}^{D4}}{ \hat{f}_i^{out}} -1   \right) \gamma_r + 1  \; ,
\end{equation}   
where $\gamma_r$ is a damping factor.\footnote{$\gamma_r = 1$ in this case. The damping factor is an optional parameter that can protect against adverse effects on the fit with the use data if necessary, see also the next section.}
A new flow matrix $F^{M5}$ can now be calculated by dividing the entries of rows $i$ in $F^{M4}$ by the corresponding correction factor $r_i$.\footnote{We avoid correcting countries with very low PR production and trade flows by demanding that $M_i > 0.005$ and $\hat{f}_i^{out} > 0.003$.} The matrix is then re-normalized equivalent to equation \ref{eq:rebal}. The statistics in table \ref{tab:fit} show that this correction in fact does improve the overall fit significantly. In particular, we increase the consistency of the flow matrix with the PR mining data without losing consistency with the fertilizer use data. We note that when comparing the fit with other models one should consider that in this step we have endogenized $M_i$.

\subsection{Simplifying the estimation}\label{sec:only}

A final consideration is whether the setup for finding the optimal weights of the goods categories can be simplified. This is important in order to gain insights about the applicability of this method to other resources or goods. While for the case of P, both the origin and the (preliminary) end use are relatively easy to determine, we can assume that for many other cases use data will not be available in comparable accuracy, since the variety in terms of applications is typically much higher than it is for P.

This means that one will typically be in a situation where a goods category weighting will have to be estimated based only on data on the origination. For the case of P this would mean that we would modify equation \ref{eq:D}, i.e., set the weight for the fertilizer use part $\alpha_U$ to 0.

We have re-estimated the model described in section \ref{sec:tie} with this setting and found that the resulting flow matrix $F^{M6}$ has a fit that is comparable with variants discussed in previous sections. 
We found that in this case a slight dampening of the correction factor in equation
\ref{eq:r} is useful by setting $\gamma_r = 0.9$. 
The fit with the mining data is very similar to that of $F^{M5}$, while the fit with the use data is only slightly worse. This might be seen as a good indication that an application of the presented model to other cases with more limited data quality should be feasible.

 \begin{figure}[tb]
  \tiny
\begin{center}
    \includegraphics[width=0.9\textwidth, trim=30 120 0 100, clip=true ]{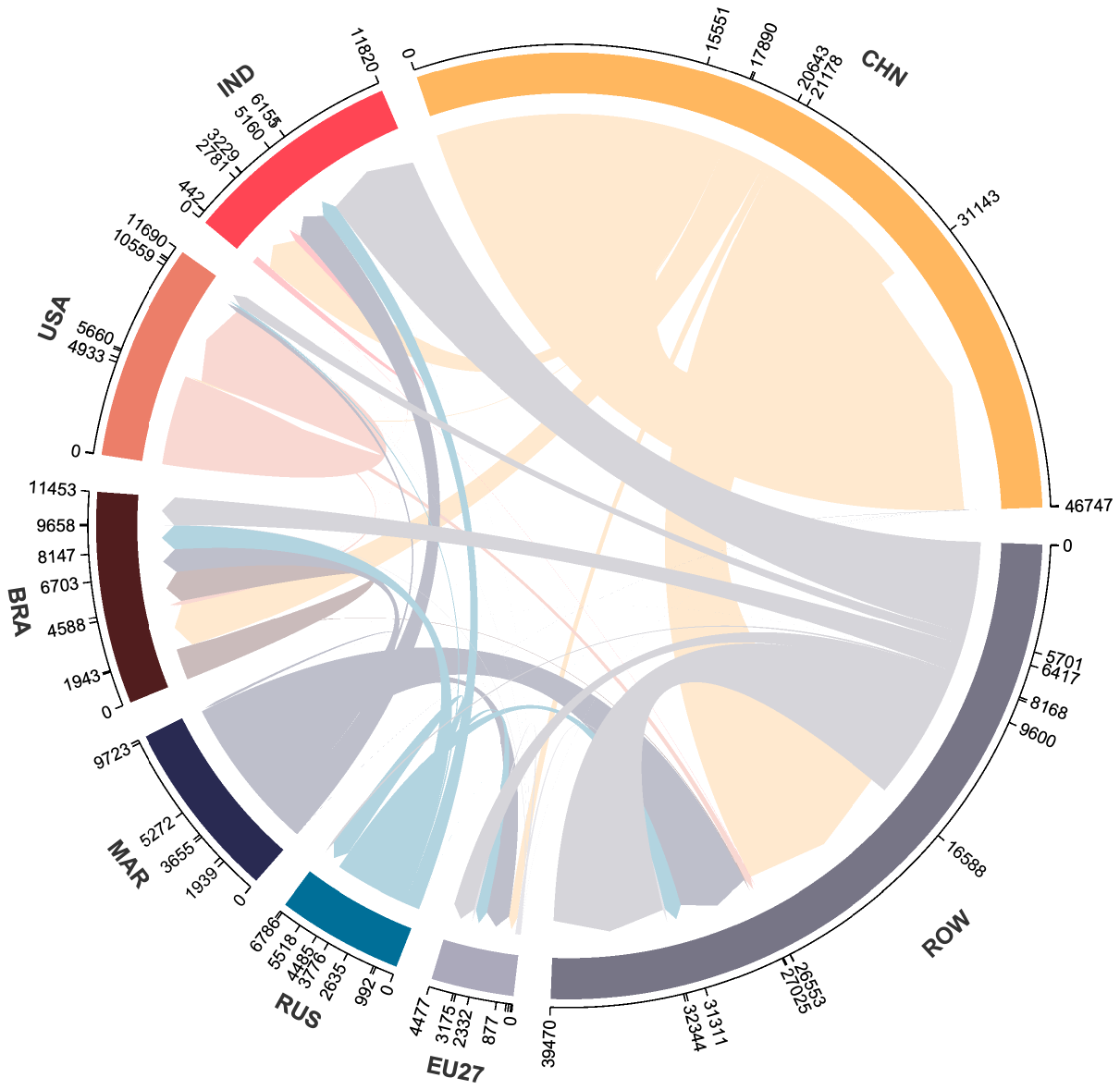}
  \end{center}
  \caption{P flow diagram based on $F^{M5}$ for 2022 for the six single countries with the largest flow as well as for the EU27 and the rest of the world (ROW), $P_2 O_5$ (thousands of tons).}\label{fig:flows}
\end{figure}

\section{Results}

\subsection{Global P flow network}

While the original trade data provides already a useful approximation of the P flows per se, it is worth noting that the ranking of the countries in terms of exports and imports differs from the ranking in terms of the material flow. This can already be observed from the aggregated data presented in the bottom of figure \ref{fig:tradesh}. For example, we observe that Belgium appears among the P exporters, but not in the panel summarizing P out-flow. Similarly, the USA do not appear in the panel describing in-flow, since from the perspective of material flow, only little P used in the USA originates from outside the country. Also, the massive changes with respect to the Chinese contribution to global phosphate flows become much clearer by looking at the material flow of P \citep[see also][for a comparison]{knight_china}.

The flow of P is visualized in more detail in figure \ref{fig:flows}. The diagram shows the flow of P based on the derived flow matrix for 2022 ($F^{M5}$). All data on the approximated flows for 2001--2022 is available for download in \href{https://cloud.tugraz.at/index.php/s/dH43JQgAfB4c65r}{spreadsheet} form. Color-coded arrows show the flow of P between as well as within countries. The cumulative amounts for each country are shown around the circle. These have been calculated by multiplying the shares of the flow by the total amount of PR mined. Note that, by construction, all flows are counted twice, once in the country of origin and once in the country of use.  

\begin{figure}[tb]
  \tiny
\begin{center}

 \includegraphics[width=0.9\textwidth, trim = 60 280 60 23, clip=true]{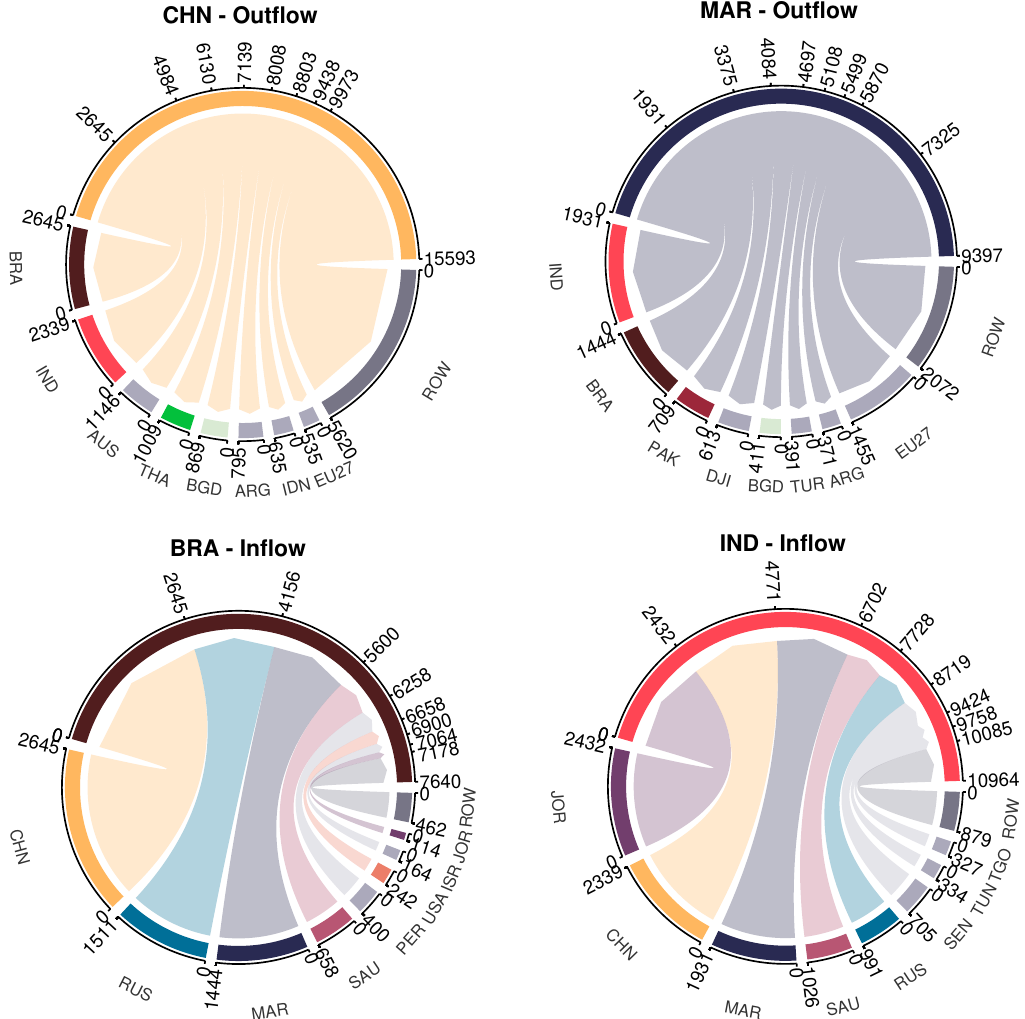}

  \end{center}
  \caption{Outflow of P for China and Morocco (top panels) and inflow of P for Brasil and India (bottom panels) for 2022, $P_2 O_5$ (thousands of tons).}\label{fig:in_out}
\end{figure}

China is the country that is involved in the largest amount of P flow. It supplies almost all the P for its own consumption, while also being a large supplier for the rest of the world. The other large supplier is Morocco, the domestic demand of which is negligible. Also, the USA mines enough PR to cover its P demand, additional inflows roughly cancel out its outflow to Brazil and some other countries. The other large consumers, India, the EU and the just mentioned Brazil differ significantly in their mix of P source countries. Brazil has some mining on its own, it completes its demand  by inflow from a mix of countries. India's mining is (relative to its size) almost negligible. It relies on inflow from Morocco, China, Peru, Russia and Saudi Arabia (here shown as part of ROW). The EU27 relies almost entirely on imports, the most important sources are Morocco and Russia. For a visualization of the flows at an earlier point in time (2001), as well as for an analysis of the fit on the country level the reader is referred to figures \ref{fig:error} and \ref{fig:in_out} in the appendix.\footnote{In general, we find that the fit for especially the largest four PR producers is reasonable. The fit of the P use data shows more variation. Here we find that or estimates for countries where the FAO data is based on official statistics tends to be rather accurate, while countries where data is based on surveys or estimates contain more outliers.}

A more detailed breakdown of the in- and outflows of the most important countries is provided in figure \ref{fig:in_out}, here flows within a country are omitted. The two top panels show the P outflow of China and Morocco, the two bottom panels show the inflow of Brazil and India.

\subsection{Dynamics of flows}

\begin{figure}[tb]
  \tiny
\begin{center}

 \includegraphics[width=0.7\textwidth, trim = 40 260 50 270, clip=true]{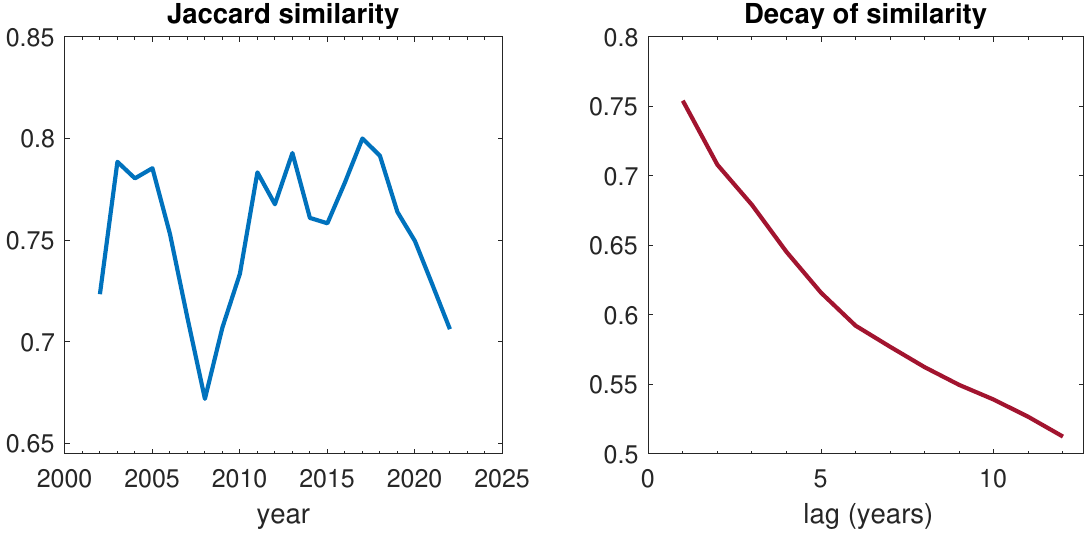}

  \end{center}
  \caption{Jaccard similarity for successive years of the flow matrix $F^{M4}$ (left panel) and decay (right panel).}\label{fig:jdist}
\end{figure}

Besides looking at the flows for a particular year, it is necessary to look at the dynamics of these flows. We measure the year-by-year changes in the flow matrix by calculating the weighted Jaccard similarity \citep[see, e.g.,][]{jac_med} between matrices as
\begin{equation}
J_{t,t-1} = \frac{\sum_{i=1}^N \sum_{j=1}^N min(F_{i,j,t},F_{i,j,t-1})}{\sum_{i=1}^N \sum_{j=1}^N max(F_{i,j,t}, F_{i,j,t-1})} \; . 
\end{equation}
Hence, a value of $J$ equal to 1 would indicate identical flow in successive years, while a value of 0 would indicate that they are are completely different.

The left panel of figure \ref{fig:jdist} shows the year-to-year changes of the flow matrix. They vary around a value of 0.75, with its lowest around the years 2007-08, which coincides with a time when P prices peaked, imports to India increased, and China's exports were volatile, see also figures \ref{fig:tradesh} and \ref{fig:trade} as well as \cite{mew_peak}. More changes and thus a drop in similarity is also observed in 2022, which coincides with the start of the Ukraine war and also higher, more volatile prices (see also appendix \ref{sec:a}).

It is still obvious that P flows show some persistence over time, and we can analyze this in more detail by looking at the decay of the Jaccard similarity over time. This means that we calculate $J$ for flow matrices which are an increasing number of years apart. The results in the right panel of figure \ref{fig:jdist} indicate that Jaccard similarity for P flows decays slowly. This gives an indication of the predictability of future P flows and the speed with which changes in P flows occur.

\subsection{Resilience}

One important application of country-level material flow analysis is the assessment of supply security and resilience. A widely used metric for this purpose is the Herfindahl-Hirschman Index (HHI), which quantifies market concentration based on the distribution of market shares \citep[see, e.g.,][]{nae_rawmat,iea24}. In the context of raw materials, the HHI is typically calculated using each country’s share in global production. For a more comprehensive view of supply security, this measure should ideally be combined with other country-level risk factors.

\begin{figure}[tb]
  \tiny
\begin{center}

 \includegraphics[width=0.7\textwidth, trim = 50 150 60 165, clip=true]{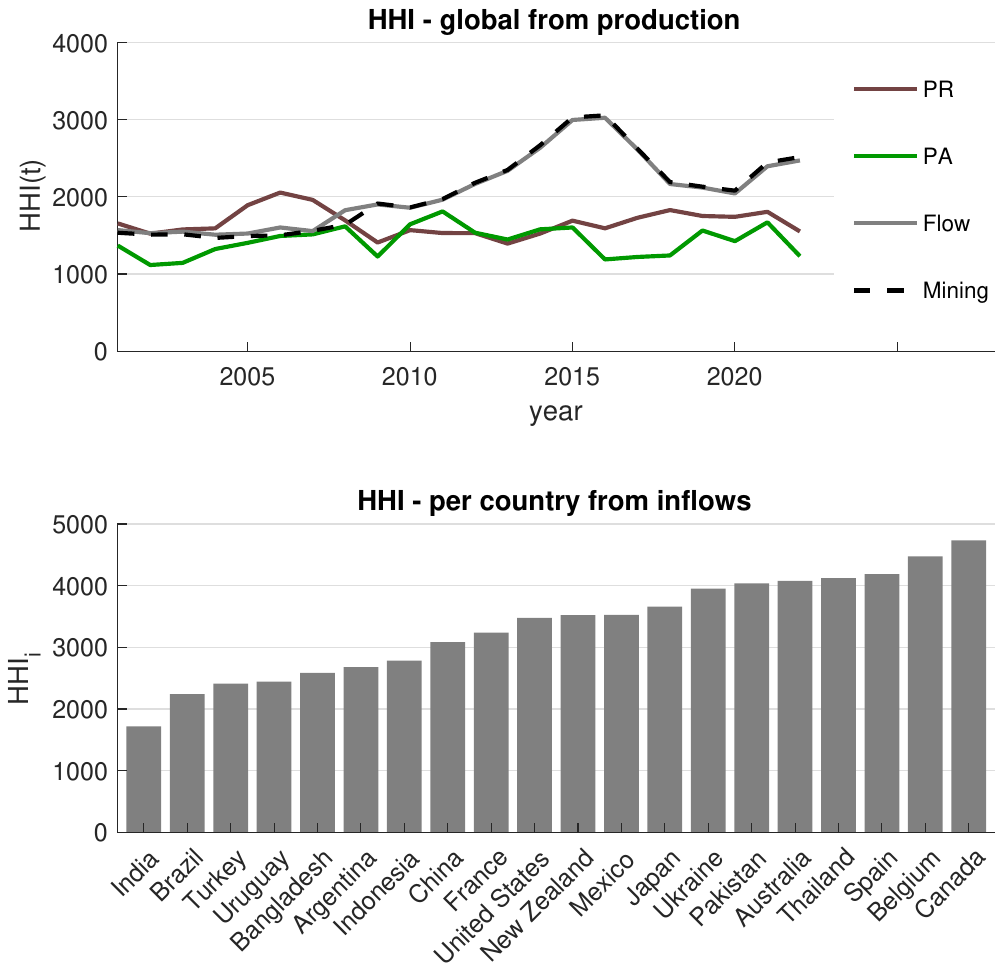}

  \end{center}
  \caption{HHI index for P over time on the global level (top panel) and Infow-HHI per country (median values for the same time period, bottom panel).}\label{fig:hhi}
\end{figure}

Let $s_i$ represent the market share of country $i$ (in percentage terms). The HHI is then calculated as

\begin{equation}\label{eq:hhi}
  HHI = \frac{1}{N} \sum_{i=1}^N s_i^2 \;  .
\end{equation}
 
The HHI ranges from just above 0 (perfect competition) to 10,000 (monopoly). Conventionally, HHI values between 1,500 and 2,500 indicate moderate market concentration, while values above 2,500 signal high concentration.\footnote{These thresholds stem from U.S. antitrust guidelines on horizontal mergers. While recent revisions have made the thresholds more stringent \citep[see][]{usj}, the older limits are still commonly used in raw material supply assessments.}

The top panel of Figure~\ref{fig:hhi} illustrates the HHI for phosphate based on mine production. Our results replicate the findings of \cite{BGR}, showing that the global HHI for phosphate has gradually increased -- from values just below 1,800 to around 2,500 -- with some year-to-year fluctuations. When using our own model-based, country-level estimates of phosphate flows, we find qualitatively similar results (partly by construction).

To provide further context, we also computed HHIs based on export market shares for phosphate rock and phosphoric acid. Up until 2011, these results closely tracked the production-based HHI. However, post-2011, they begin to diverge. The rising HHI for mine production reflects the sharp increase in China’s phosphate rock production between 2010 and 2016. Interestingly, this rise did not correspond to a similar increase in China’s share of P-related exports. One plausible explanation is that the reported Chinese mining figures during that period may have been slightly overstated. Supporting this, we also observe a discrepancy between global mining and fertilizer use data during the same time frame, likely reflecting the same underlying issue.

More importantly, our estimated P flow data allows us to calculate a country-specific version of the HHI, which we refer to as the Inflow-HHI. This metric captures the concentration of phosphorus supply sources for each importing country. To compute it, we normalize each country’s total phosphorus inflow (based on $F^{M5}$) to 100\% and calculate the HHI based on the shares of imports from individual source countries, using the same formula as above.

The bottom panel of Figure~\ref{fig:hhi} presents Inflow-HHI values for the 20 largest importing countries. Two key insights emerge:
First, country-level HHIs are generally much higher than the global HHI, indicating greater concentration and thus higher vulnerability for individual countries.
Second, there is significant variation among countries. For example, major importers like India and Brazil have Inflow-HHIs around 2,000, suggesting moderate concentration. In contrast, countries such as Spain, Belgium, and Canada exhibit HHIs exceeding 4,000, indicating high dependency on a limited number of suppliers.

It is important to note that this Inflow-HHI reflects only the concentration of external supply sources. For countries with significant domestic phosphate production -- such as the United States and China -- the Inflow-HHI should be interpreted in the context of their overall import dependence.

\section{Conclusions}

This study demonstrates that trade data, when combined with other relevant data sources, can be effectively used to approximate the flow of mineral resources -- even with only limited prior knowledge about material concentrations. Specifically, our approach provides a meaningful representation of global phosphorus flows -- encompassing phosphate rock, fertilizers, and related products -- prior to biomass production. This offers a valuable foundation for analyzing vulnerabilities of country-level supply relationships, particularly relevant with respect to food security.

A key contribution of this work is the translation of nominal bilateral trade data into material flows of phosphorus, which enhances the accuracy of supply chain assessments. This is achieved through a top-down approach, involving the indirect estimation of material concentrations in traded goods, as well as by algorithms that correct data on net exports such that they can be interpreted as flows. Our method yields insights into:
\begin{itemize}\setlength{\itemsep}{-4pt}
    \item the origin of phosphorus flows,
    \item their destinations and approximate material content,
    \item the resilience of national supply structures, and
    \item the complex web of global dependencies in phosphorus trade.
\end{itemize}
\noindent We show that the resulting flows can be used for a significantly improved analysis of the resilience of phosphate supply, exemplified by the calculation of Inflow-HHIs on the country level.

Further, the integration of multiple data sources also allows for the identification of inconsistencies between reported data and model-based expectations, offering a way to validate and improve existing datasets.

Finally, the approach developed here is generalizable. While demonstrated using phosphorus, it can be extended to other critical raw materials -- such as sulfur, nitrogen, or potassium -- that are essential for agriculture and industry.

\newpage
\subsection*{Author contributions}
Conceptualization, all authors; Methodology, M.R.; Formal Analysis, M.R.; Data Curation, M.R.; Writing – Original Draft Preparation, all authors; Writing – Review \& Editing, all authors.; Visualization, M.R.

\subsection*{Acknowledgments} The authors thank F.-W. Wellmer, R.W. Scholz, M. Mew, P. Klimek, P. Hirner and S. Havlin for discussions of this work. 

\subsection*{Declaration of interest}
Funding: GS received partial funding from the Global Phosphorus Institute.\\
\noindent Other: none.

\subsection*{Data availability}
The flow matrices are available for download at \href{https://cloud.tugraz.at/index.php/s/dH43JQgAfB4c65r}{\small{https://cloud.tugraz.at/index.php/s/dH43JQgAfB4c65r}}. Source code is available upon request. All used data is publicly available.
\clearpage

\newpage
\section*{Appendix}
\appendix

\section{Details on data}\label{sec:a}

 \begin{figure}[htb]
  \tiny
\begin{center}
    \includegraphics[width=0.8\textwidth, trim= 5 50 10 100, clip=true]{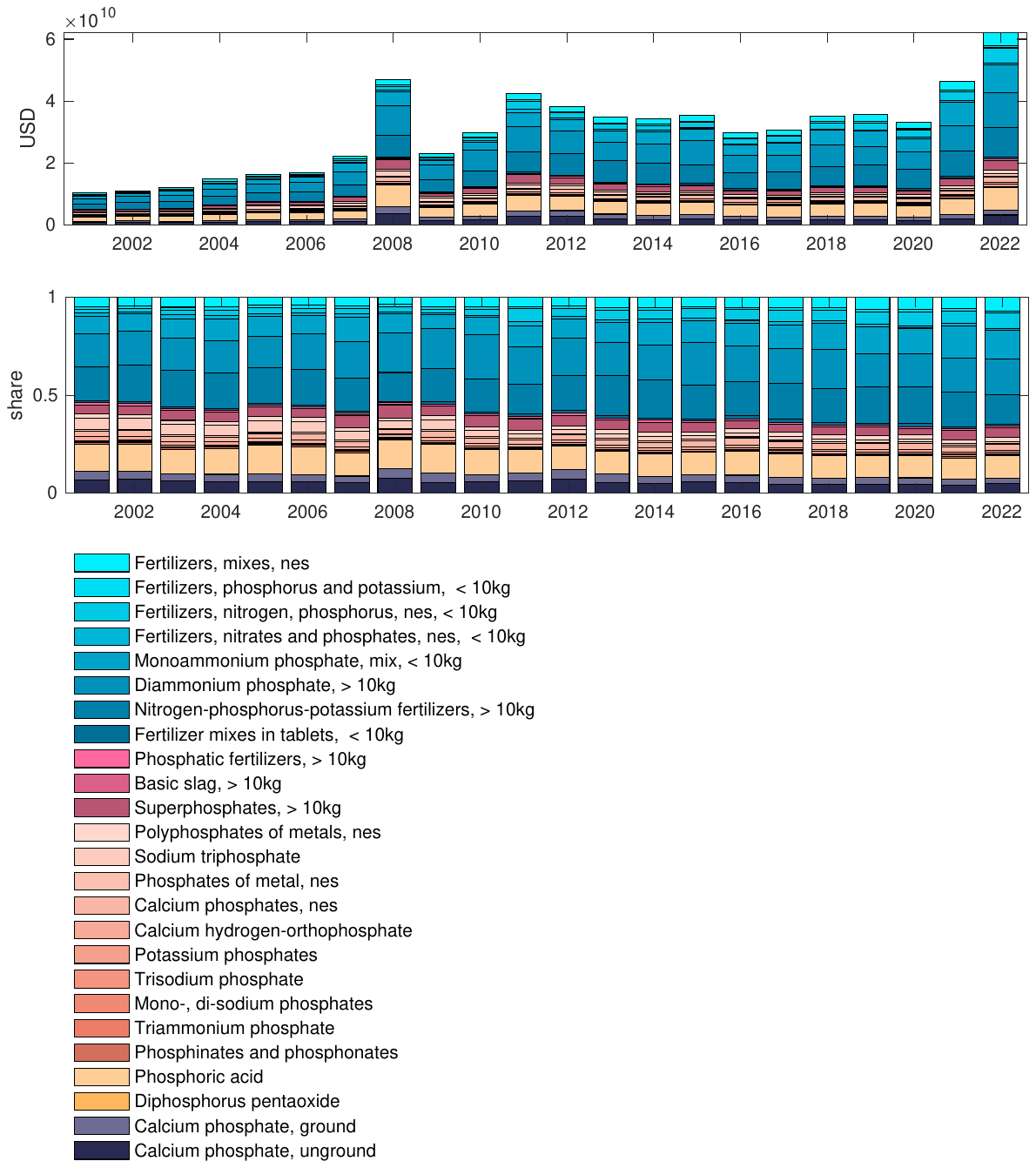}
  \end{center}
  \caption{Trade in P-related goods, 2001--2022. The top panel shows the USD amounts of trade in P-related goods, dis-aggregated into the 25 6-digit HS categories. The bottom panel shows the fractions of these 25 goods categories over time.}\label{fig:trade}
\end{figure}

\clearpage
\section{Details on methods}

\begin{figure}[htb]
  \tiny
\begin{center}
    \includegraphics[width=0.8\textwidth, trim= 39 220 100 60, clip=true]{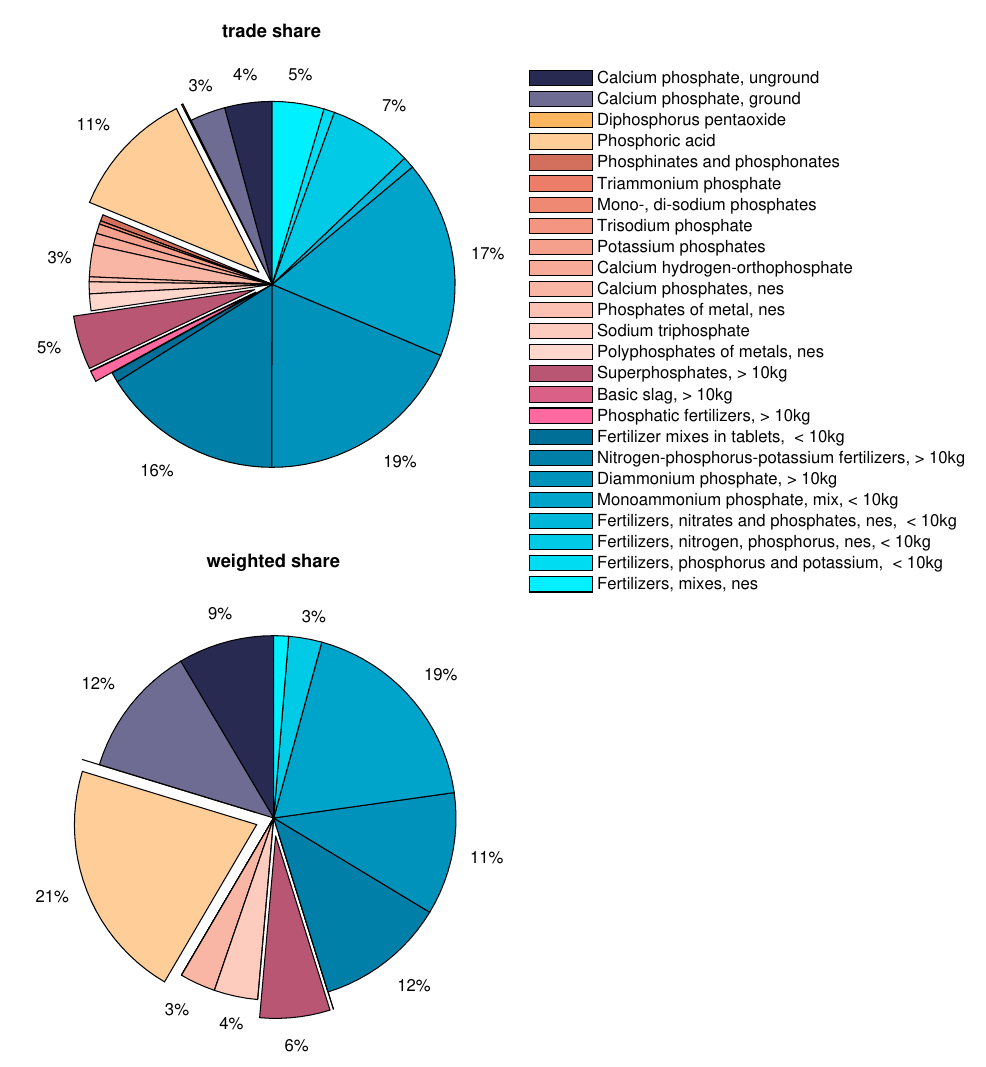}
  \end{center}
  \caption{Contributions of goods categories. The top chart shows the shares of the different goods categories in terms of USD of net trade in 2021. The bottom chart shows the shares based on weighted net trade as used in the estimation of the flow matrix $F^{M5}$. The legend shows the 25 dis-aggregated goods categories. In each panel percentage shares are given for the goods categories (labels $\leq 1\%$ omitted).}\label{fig:pie}
\end{figure}

\begin{figure}[htb]
  \tiny
\begin{center}
    \includegraphics[width=0.8\textwidth, trim= 30 200 30 180, clip=true]{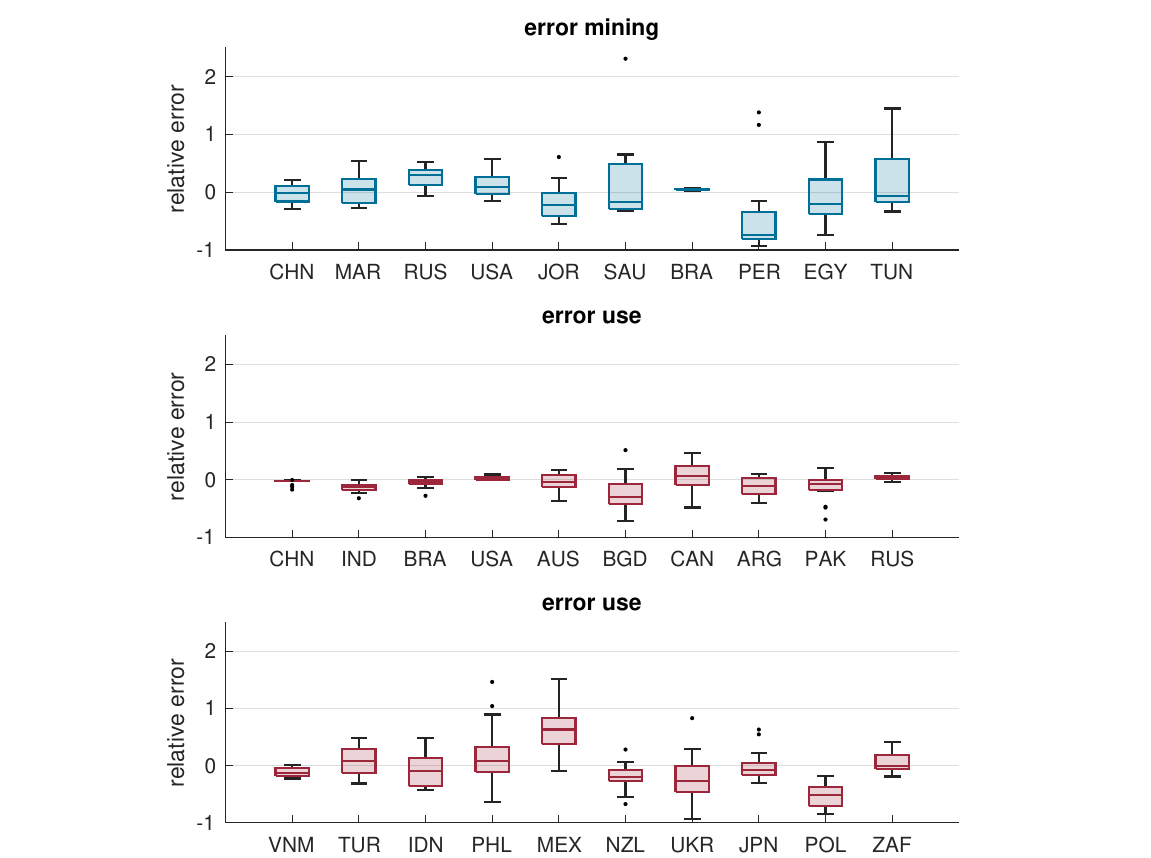}
  \end{center}
  \caption{Box-plots of the deviations of the estimated PR mining (top panel) and P use (middle and bottom panel) relative to the empirical data (2001--2022), i.e., $M$ and $U$ for the 10 largest producer and 20 largest P fertilizer user countries respectively based on $F^{M4}$.}\label{fig:error}
\end{figure}

 \begin{figure}[htb]
  \tiny
\begin{center}
    \includegraphics[width=0.9\textwidth, trim=40 120 20 110, clip=true ]{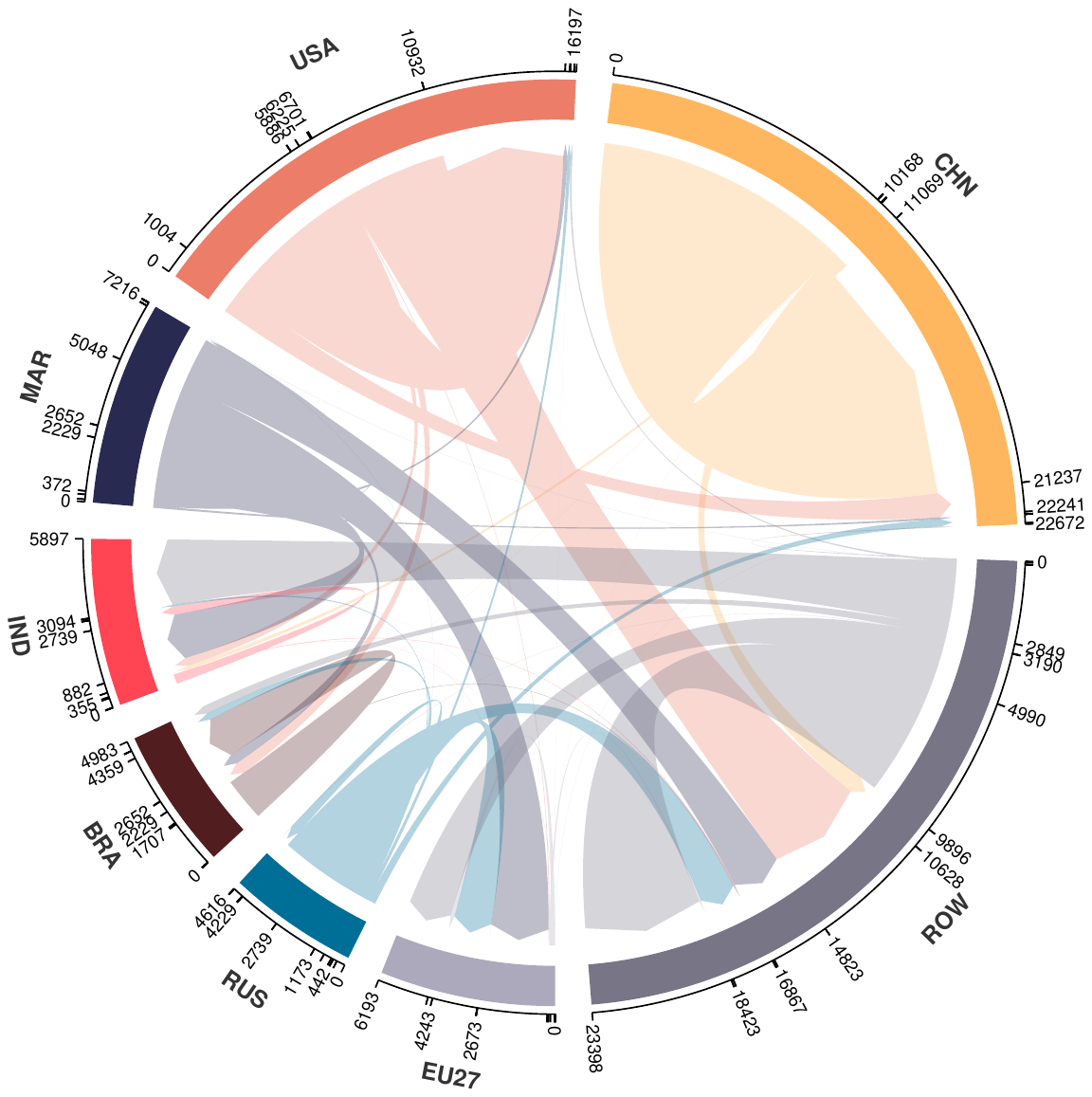}
  \end{center}
  \caption{P flow diagram based on $F^{M5}$ for 2001 for the six single countries with the largest flow as well as for the EU27 and the rest of the world (ROW), $P_2 O_5$ (thousands of tons).}\label{fig:flows01}
\end{figure}

\clearpage



\bibliographystyle{elsarticle-harv} 
\bibliography{network2}

\end{document}